\documentclass[conference]{IEEEtran}
\IEEEoverridecommandlockouts

\usepackage{cite}
\usepackage{amsmath,amssymb}
\usepackage{newtxtext,newtxmath}  
\usepackage{graphicx}
\usepackage{textcomp}
\usepackage{url}
\usepackage{xcolor}
\usepackage{fancyhdr}
\usepackage{balance}

\fancypagestyle{ieeefirst}{%
  \fancyhf{}%
  \fancyfoot[L]{\footnotesize 979-8-3195-2794-3/26/\$31.00~\textcopyright2026 IEEE}}

\graphicspath{{figures/}}

\newcommand{\rev}[1]{#1}          

\begin{document}

\title{The CosmicWatch Desktop Muon Detector (v3X):\\
Prototype Design and Adaptation for Spaceflight}

\author{%
\IEEEauthorblockN{Anvay Ajmera}
\IEEEauthorblockA{%
AxLab, Dept.\ of Physics and Astronomy\\
University of Delaware\\
Newark, DE, USA\\
ORCID: 0009-0009-4149-4178}
\and
\IEEEauthorblockN{Spencer N. Axani}
\IEEEauthorblockA{%
Dept.\ of Physics and Astronomy\\
University of Delaware\\
Newark, DE, USA\\
ORCID: 0000-0001-8866-3826}
}

\maketitle
\thispagestyle{ieeefirst}

\begin{abstract}
The CosmicWatch Desktop Muon Detector (v3X) is a compact and low-cost scintillator-based
instrument designed to detect ionizing radiation, including cosmic-ray muons. Built around a
plastic scintillator, a silicon photomultiplier (SiPM), and a dual-core Raspberry Pi Pico
microcontroller, the v3X achieves a 400\,$\mu$s dead time per event, sustained event logging
at 700\,Hz, and two-detector coincidence triggering for more accurate counts. This paper
first summarizes the design, firmware, and demonstrated performance of the v3X prototype. It
then outlines the design changes needed to prepare the detector for operation aboard a
satellite, including a passive copper thermal strap that conducts heat from the
microcontroller to the spacecraft chassis, a dedicated flight enclosure, material changes to
meet outgassing standards, and a Controller Area Network (CAN) bus interface to connect the
detector to the host spacecraft. The flight unit is targeted for launch via Oligo Space.
\end{abstract}

\section{Introduction}

CosmicWatch \cite{Axani2018,Axani2019} is a University of Delaware (UD)-led outreach and
instrumentation program built to introduce students to the intersection between particle
physics and electronics by building particle detectors and performing physics measurements.
Since its origin at the Massachusetts Institute of Technology (MIT) and the Polish National
Center for Nuclear Research in 2017, the program has grown into a strong platform used for
education, calibration studies, and spacecraft instrumentation, while also teaching other
practical skills such as soldering, microcontroller programming, and data analysis.

Over the past year, a new generation of the detector, the v3X \cite{Axani2025}, has been
developed with substantial improvements over the earlier v1 and v2 models, with the total
component cost staying near \$100. This paper reviews the design and demonstrated
performance of the v3X \cite{Axani2025}, and presents the in-progress modifications on an
updated prototype v3X to adapt it for long-duration operation aboard a satellite, such as
passive thermal management, a flight enclosure, and a CAN bus communication interface, all
with \rev{supporting flight-interface changes}. The flight unit is planned for launch via
Oligo Space.

\begin{figure}[!htb]
\centering
\includegraphics[width=0.68\columnwidth]{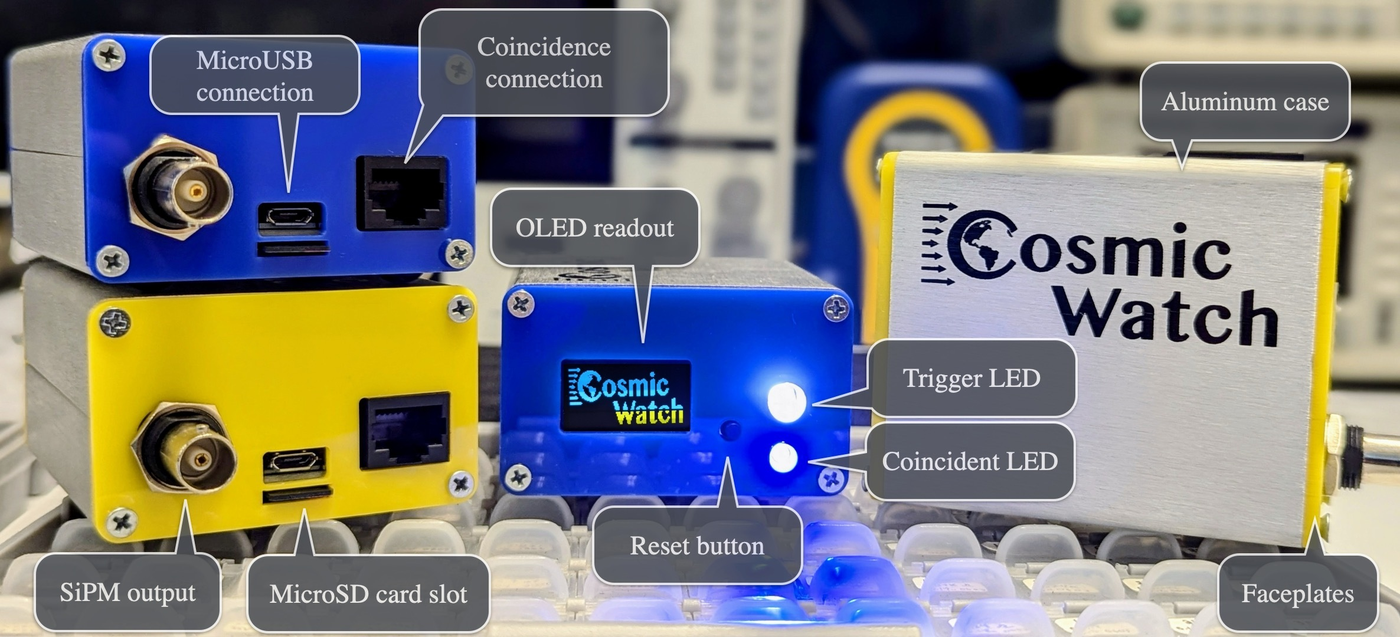}
\caption{An array of CosmicWatch Desktop Muon Detectors (v3X), showing the OLED readout, SiPM analog
output (BNC), coincidence port (RJ45), microSD slot, and aluminum enclosure. Reproduced from
\cite{Axani2025}.}
\label{fig:array}
\end{figure}

\section{The CosmicWatch v3X Detector}

\subsection{Principle of Operation}

When a charged particle passes through the detector's 5$\times$5$\times$1\,cm$^{3}$
polystyrene-based plastic scintillator, it excites the scintillant, which de-excites by
emitting photons, resulting in approximately 10,000 photons/MeV of deposited energy. A
fraction of this light is collected by a 6$\times$6\,mm$^{2}$ silicon photomultiplier (SiPM)
optically coupled to the scintillator and operated at 30\,V reverse bias, which corresponds
to a 5.5\,V overvoltage and a photon detection efficiency of about 43\% at 420\,nm. The SiPM
can then convert the collected photons into an electrical pulse, where the amplitude is
proportional to the light intensity, and therefore to the energy deposited in the
scintillator.

From here, the SiPM signal is amplified and routed to a comparator, which timestamps the
event, and to a peak detector, which holds a constant fraction of the pulse amplitude for
digitization by the microcontroller's 12-bit ADC. The raw SiPM pulse is also available on a
rear-panel BNC connector for external readout with nanosecond timing resolution. To suppress
radiogenic backgrounds, two detectors can be operated in coincidence mode; in this, events
triggering both units within a 2.3\,$\mu$s window are flagged as coincident, which
effectively selects through-going cosmic-ray muons and rejects uncorrelated gamma and beta
backgrounds.

The geometry of the detector pair also directly influences the measured muon count. For
instance, having the two detectors flat side-by-side leaves both scintillators open to
nearly the full overhead sky to maximize the coincidence rate, while stacking them
vertically with a gap in between them shortens the accepted solid angle so that only muons
arriving from a particular direction can trigger both units; when adapted, this allows for
both directional and angular-distribution measurements \cite{Axani2025}.

\begin{figure}[!htb]
\centering
\includegraphics[width=0.96\columnwidth]{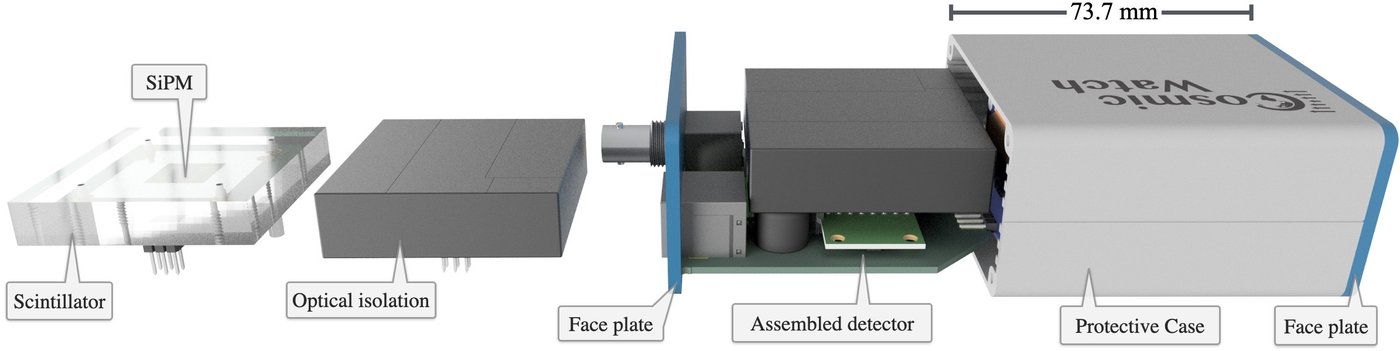}
\caption{The full assembly of the CosmicWatch v3X detector: scintillator, optical isolation, SiPM
PCB, main PCB, aluminum protective case, and laser-cut faceplates. Reproduced from
\cite{Axani2025}.}
\label{fig:assembly}
\end{figure}

\subsection{Electronics Design}

All onboard electronics are powered from the 5\,V USB input on the Raspberry Pi Pico
(RP2040) microcontroller \cite{RP2040,Pico}. A MAX5026 DC-DC booster generates the
$\sim$30.0\,V SiPM bias with a measured output ripple of $\leq$10\,mV, along with a
precision 2.5\,V reference that biases the analog chain and the ADC. The amplification stage
uses an AC-coupled TPH2502 operational amplifier in a non-inverting configuration with a
voltage gain of 13.5\,V/V and approximately 10\,MHz of bandwidth. A second operational
amplifier implements the comparator, whose trigger threshold is set by an RC-filtered PWM
output of the microcontroller and is user-configurable down to 4\,mV.

Digital processing is handled by the Pico's dual-core ARM Cortex-M0+ processor. The board
also hosts a BMP280 pressure and temperature sensor, an MPU-6050 accelerometer and
gyroscope, a microSD card socket, an OLED display, and a buzzer. An RJ45 coincidence port
carries power and trigger signals between paired detectors, allowing a single USB source to
power multiple daisy-chained units.

\subsection{Firmware and Data Acquisition}

The firmware, written in C++, heavily utilizes the Pico's dual-core architecture, as Core 0
runs a tight polling loop dedicated to time-critical acquisition, with comparator
monitoring, coincidence polling, ADC sampling, timestamping, and dead-time accounting, while
Core 1 manages the OLED and environmental sensors and handles microSD logging, holding
events in memory and writing them to the card in batches. This separation sustains event
logging at 700\,Hz to both the serial port and microSD card simultaneously (approximately
7,000\,Hz with serial output disabled), with an average dead time per event of about
400\,$\mu$s.

Each recorded event contains the event number, timestamp, coincidence flag, 12-bit ADC
value, reconstructed SiPM pulse amplitude, cumulative dead time, temperature, barometric
pressure, linear acceleration, and angular velocity, saved as plain text with one line per
event and a fixed set of columns. Data can be viewed on the OLED, logged to the microSD
card, or streamed over USB, where a host-side Python script adds NTP-synchronized
timestamps. A configuration file on the microSD card has user-adjustable settings, including
the trigger threshold and detector name.

\subsection{v3X Performance}

\begin{table}[!t]
\caption{Selected specifications of CosmicWatch v3X \cite{Axani2025}.}
\label{tab:specs}
\centering
\renewcommand{\arraystretch}{1.15}
\begin{tabular}{@{}ll@{}}
\hline
\textbf{Specification} & \textbf{v3X} \\
\hline

Processor & RP2040 Cortex-M0+ \\

Cores / clock & 2 / 133\,MHz \\

ADC resolution & 12-bit \\

Max. event rate & 700\,Hz \\

Dead time per event & 400\,$\mu$s \\

Noise level & 0.1\,mV RMS \\

Min. trigger threshold & 4\,mV \\

Accidental coinc. rate & 4.5$\times$$10^{-5}$\,Hz \\

Accel./gyro., buzzer & Yes \\

Approximate cost & \$100 \\

\hline
\end{tabular}
\end{table}

Table~\ref{tab:specs} summarizes selected hardware and performance specifications of the
v3X, including its processor, acquisition rate, dead time, noise level, trigger threshold,
coincidence performance, onboard sensing, and component cost.

\section{Example Measurement: High-Altitude Balloon}

\begin{figure}[!htb]
\centering
\includegraphics[width=0.92\columnwidth]{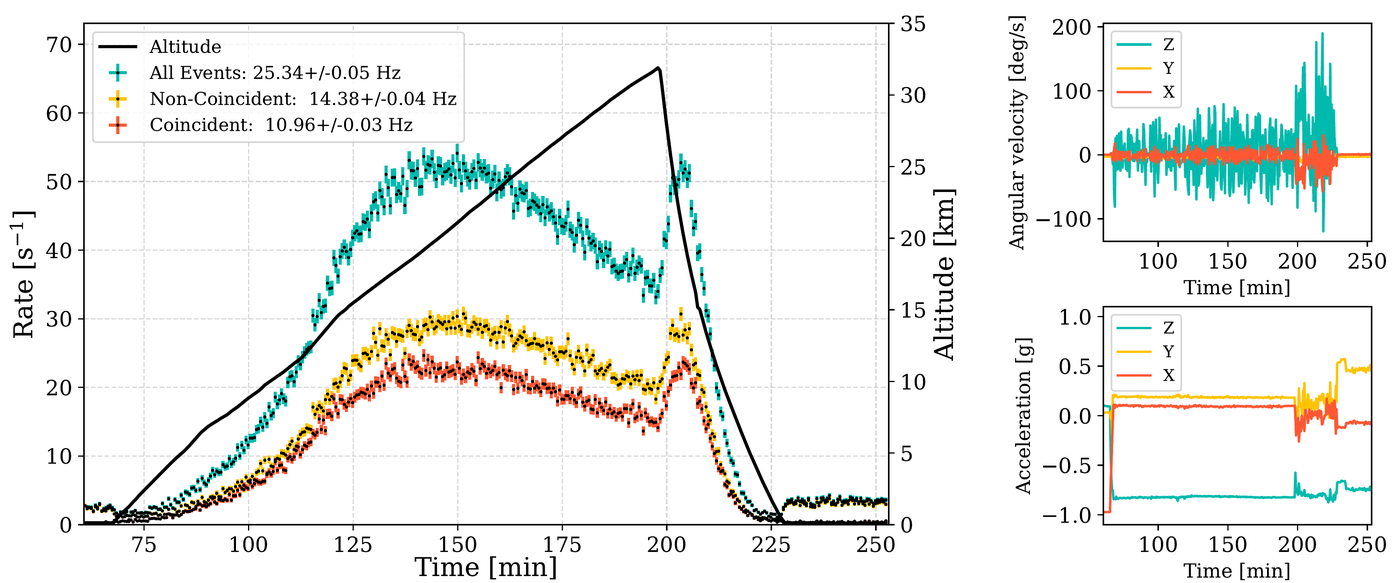}
\caption{The left shows event rate versus time during a high-altitude balloon flight, with the
altitude calculated from the onboard pressure sensor shown on the right axis, where the rate
peaks at the Regener-Pfotzer maximum near 20\,km. The right shows angular velocity (top) and
linear acceleration (bottom) recorded by the onboard IMU during the flight. Reproduced from
\cite{Axani2025}.}
\label{fig:balloon}
\end{figure}

Two detectors were flown on a high-altitude balloon to 31\,km, recording continuously to
their microSD cards while running off a single 5,000\,mAh battery pack. The measured event
rate as a function of altitude (Fig.~\ref{fig:balloon}) reproduces the characteristic
atmospheric profile: the rate climbs to a broad maximum near 20\,km, the Regener-Pfotzer
maximum \cite{Regener}, where secondary cosmic-ray production and attenuation balance, then
decreases at higher altitude.

Both detectors were operated in coincidence mode, in which events triggering both units
within a 2.3\,$\mu$s window are flagged as coincident and uncorrelated backgrounds
contribute only a flat accidental rate

\begin{equation*}
R_{\mathrm{acc}} = 2N_1N_2\tau \approx 4.5\times10^{-5}\,\mathrm{Hz}
\end{equation*}

where $N_1$ and $N_2$ are the individual detector trigger rates and $\tau = 2.3\,\mu$s is
the coincidence window \cite{Axani2025}. The energy a muon loses crossing a thin absorber
follows a Landau distribution, which can be approximated by the analytic Moyal form
\cite{Axani2025}:

\begin{equation}
f(x;\mu,c) = \frac{1}{\sqrt{2\pi}\,c}\,
\exp\!\left[-\tfrac{1}{2}\left(z + e^{-z}\right)\right]
\label{eq:moyal}
\end{equation}

where $z = (x - \mu)/c$, $\mu$ is the most probable pulse height, and $c$ is the width
parameter; at sea level the fitted peak ($\mu \approx 21$\,mV) matches the $\approx$2\,MeV
that a minimum-ionizing muon deposits in 1\,cm of plastic scintillator, so coincidence
triggering combined with a simple pulse-height cut cleanly isolates cosmic-ray muons from
the radiogenic background \cite{Axani2025}.

The onboard accelerometer and gyroscope tracked payload orientation and rotation throughout
the flight: the payload hung sideways at launch, re-oriented vertically shortly after
liftoff, and rotated continuously about the zenith. The detectors, housed in a polystyrene
enclosure with a Mylar heat blanket, stayed between 45\,$^{\circ}$C (during lift-off and
landing) and 23\,$^{\circ}$C (at apogee) \cite{Axani2025}.

\section{Modifications for Space Flight}

A prototype pair of v3X detectors is being adapted to fly as a hosted payload mounted to a
satellite. This mission introduces constraints that the detector was not designed for, such
as heat only being able to leave the system by conduction and radiation, the assembly
needing to survive launch vibration, the electrical interface to the spacecraft being
limited to four wires, and the instrument having to operate unattended for roughly three
years with no possibility of troubleshooting. This section describes the \rev{four major
modifications, including vacuum-compatible material selection, }passive thermal management,
a customized flight enclosure, and a CAN bus interface.

\subsection{\rev{Vacuum-Compatible Material Selection}}

\rev{The satellite-adapted v3X detector pair will pursue a 510\,km sun-synchronous orbit
mission, which imposes contamination and material requirements implemented by Oligo Space
and SpaceX Rideshare. Given that the payload materials must comply with NASA-STD-8719.24 and
satisfy ASTM E595 limits of less than 1.0\% total mass loss and less than 0.1\% collected
volatile condensable material, the original v3X pair's scintillator and optical isolation
materials fail these requirements.}

\rev{To solve this, the prototype flight detector now uses a polyvinyl toluene (PVT)-based
plastic scintillator, which converts energy deposited by ionizing radiation into optical
photons. Optical isolation is provided by 3M 425 aluminum foil tape, part number 425-1-5,
with an aluminum backing and acrylic adhesive, with a listed operating range of
$-$54\,$^{\circ}$C to +149\,$^{\circ}$C. These materials, along with the updated prototype,
pass the outgassing review and qualification process to continue the mission
\cite{ICD,DigiKey}.}

\rev{Additionally, standard commercial PCB assemblies are not inherently certified to ASTM
E595, since the laminate, solder mask, solder, connectors, wire insulation, adhesives,
labels, and component packaging are typically all uncertified. So, before delivery, flux
residues will be removed, and the assembled payload is scheduled for a 24\,h bakeout at
80\,$^{\circ}$C and below $10^{-5}$\,Torr, followed by contamination inspection \cite{ICD}.}

\subsection{Passive Thermal Management}

On the ground, air convection cools the detector electronics, while in orbit, the lack of
air means heat must be transferred through conduction to the spacecraft structure and then
dissipated by radiation. The dominant heat source on the main PCB is the RP2040
microcontroller. Therefore, to provide a low-resistance conductive path, a passive thermal
strap was designed consisting of a flexible copper braid sandwiched between two small copper
plates (approximately 8$\times$8\,mm$^{2}$, in order to match the RP2040 size). One end of
the strap sits on top of the RP2040, while the braid's far end connects to the aluminum
structure of the satellite, which acts as the thermal reservoir that carries the heat away.

\rev{The host spacecraft provides passive thermal management but no dedicated active
thermal-control service. Its mounting interface is specified to remain between
$-$20\,$^{\circ}$C and +80\,$^{\circ}$C during nominal operation, making the back plate and
thermal strap the payload's principal conductive path to the spacecraft structure
\cite{ICD}.}

\begin{figure}[!htb]
\centering
\includegraphics[width=0.72\columnwidth]{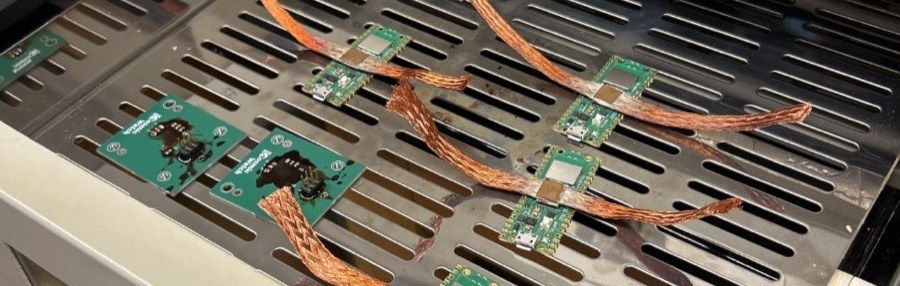}
\caption{\rev{Improved thermal-strap assemblies after the 150\,$^{\circ}$C / 1\,h cure, with the
copper braid centered between the plates in the flight orientation.}}
\label{fig:straps}
\end{figure}

The first prototype straps were assembled using only EPO-TEK 353ND \cite{EpoTek}, which is a
two-component, high-temperature epoxy developed for semiconductor and fiber-optic assembly.
The epoxy is mixed 10:1 by weight, has a pot life of about three hours, and, most
importantly for this specific application, passes the NASA low-outgassing standard (ASTM
E595) when properly cured, since an adhesive that slowly released trapped gases in vacuum
could contaminate nearby electronics and optics on the spacecraft. It also holds strong at
high temperature, with a glass transition temperature of at least 90\,$^{\circ}$C and a
degradation temperature of 412\,$^{\circ}$C, both well above anything expected here.
\rev{The first assemblies used the 80\,$^{\circ}$C / 30 min minimum schedule; the improved
straps in Fig.~\ref{fig:straps} were cured at 150\,$^{\circ}$C for 1\,h in order to provide
a more complete cure, and are now the selected fabrication profile.}

\begin{figure}[!htb]
\centering
\includegraphics[width=0.64\columnwidth]{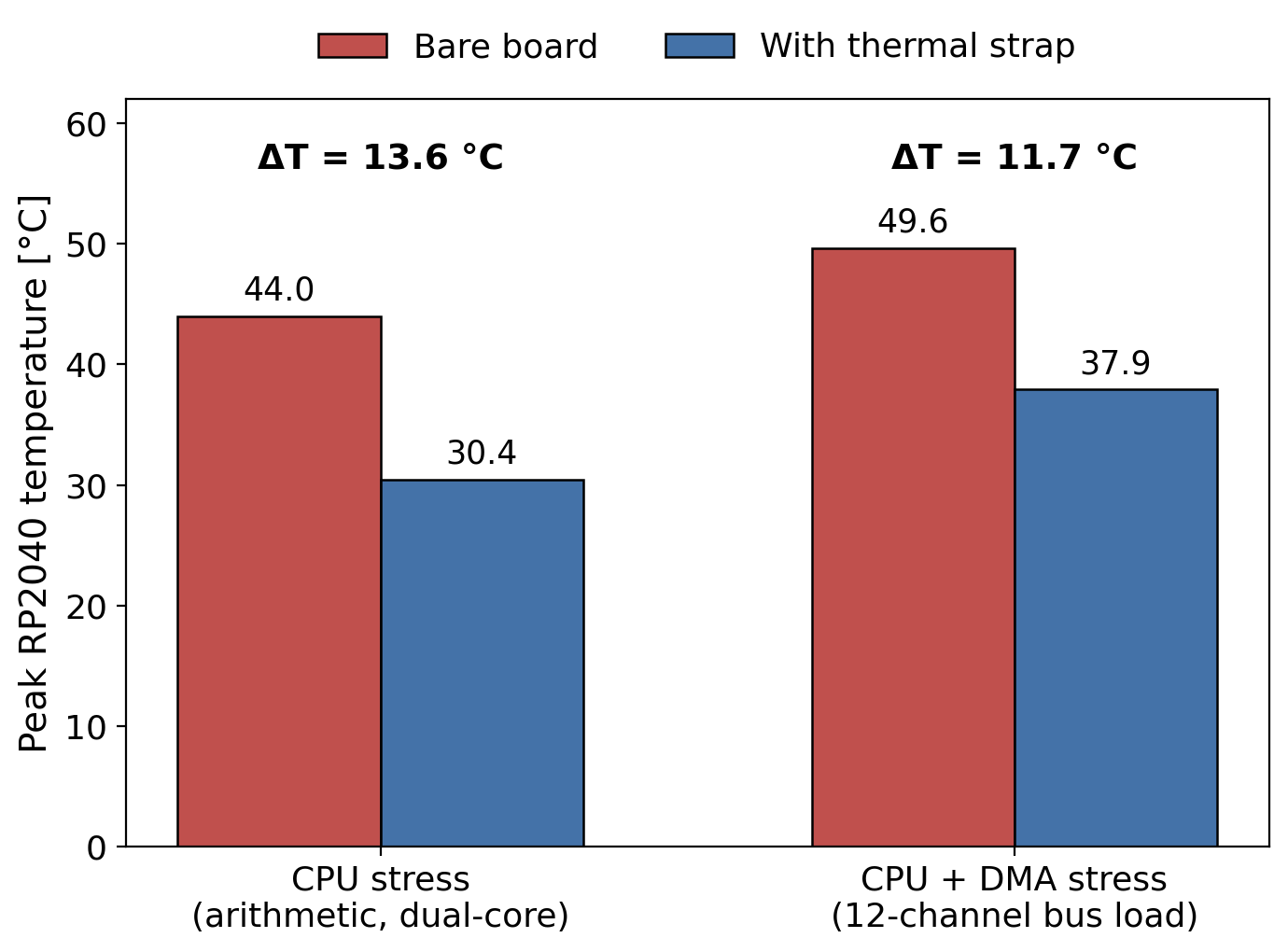}
\caption{Peak RP2040 temperature during 15-minute stress tests with and without the copper thermal
strap, for a dual-core arithmetic load and a combined CPU + 12-channel DMA load.}
\label{fig:stress}
\end{figure}

To measure how much the strap truly helps, a stress-test firmware was written that pushes
the Pico to a repeatable high-heat load. The program reads the RP2040's built-in temperature
sensor, an analog sensor on the chip itself that the Pico digitizes through its ADC, once
per second through an idle baseline period, a 15-minute load phase, and a cooldown. The
first version of the test keeps both CPU cores running arithmetic-heavy loops with the
system clock raised to 300\,MHz at elevated core voltage; the second additionally claims 12
DMA channels that continuously copy data between memory locations, so that the chip's
internal bus is loaded as well. On a bare board, the chip peaked at 43.99\,$^{\circ}$C under
the CPU-only load and 49.61\,$^{\circ}$C with CPU and DMA combined; with the thermal strap
mounted, the same tests peaked at 30.42\,$^{\circ}$C and 37.91\,$^{\circ}$C, which are
reductions of 13.6\,$^{\circ}$C and 11.7\,$^{\circ}$C respectively (Fig.~\ref{fig:stress}).

All of these tests were run at room temperature, with each board starting from an idle
temperature of roughly 20--22\,$^{\circ}$C. Here, ordinary air convection helps cool both
configurations, and the strap ends in a modest heat sink, so the measured reductions
understate the benefit expected in vacuum, where the strap provides the only conductive path
to the chassis. The next steps are computer heat-flow simulations, thermal-camera imaging of
the powered assembly, and testing in a thermal-vacuum (TVAC) chamber, which reproduces the
temperature swings of orbit.

The epoxy-only assembly, however, revealed an important weakness at the plate-braid-plate
joint. The porous copper braid draws liquid epoxy into its empty spaces via capillary action
and leaves too little adhesive at the copper-to-copper bond; the resulting joint is
mechanically weaker, has poor thermal contact, and can result in the braid snapping off the
plate. The fix is to join the copper stack with solder instead: it wets the copper surfaces,
fills the air gaps within the braid, and creates a continuous metallic joint with higher
thermal conductivity than an epoxy bond. The epoxy is now solely designated for the joint
between the finished strap and the RP2040 package, where soldering temperatures would damage
the chip. The measurements above were taken with epoxy-only straps, so the solder-joined
version is expected to perform at least as well, and the TVAC results should show a larger
benefit still once convection is removed.

\subsection{Flight Enclosure}

\begin{figure}[!htb]
\centering
\includegraphics[width=0.88\columnwidth]{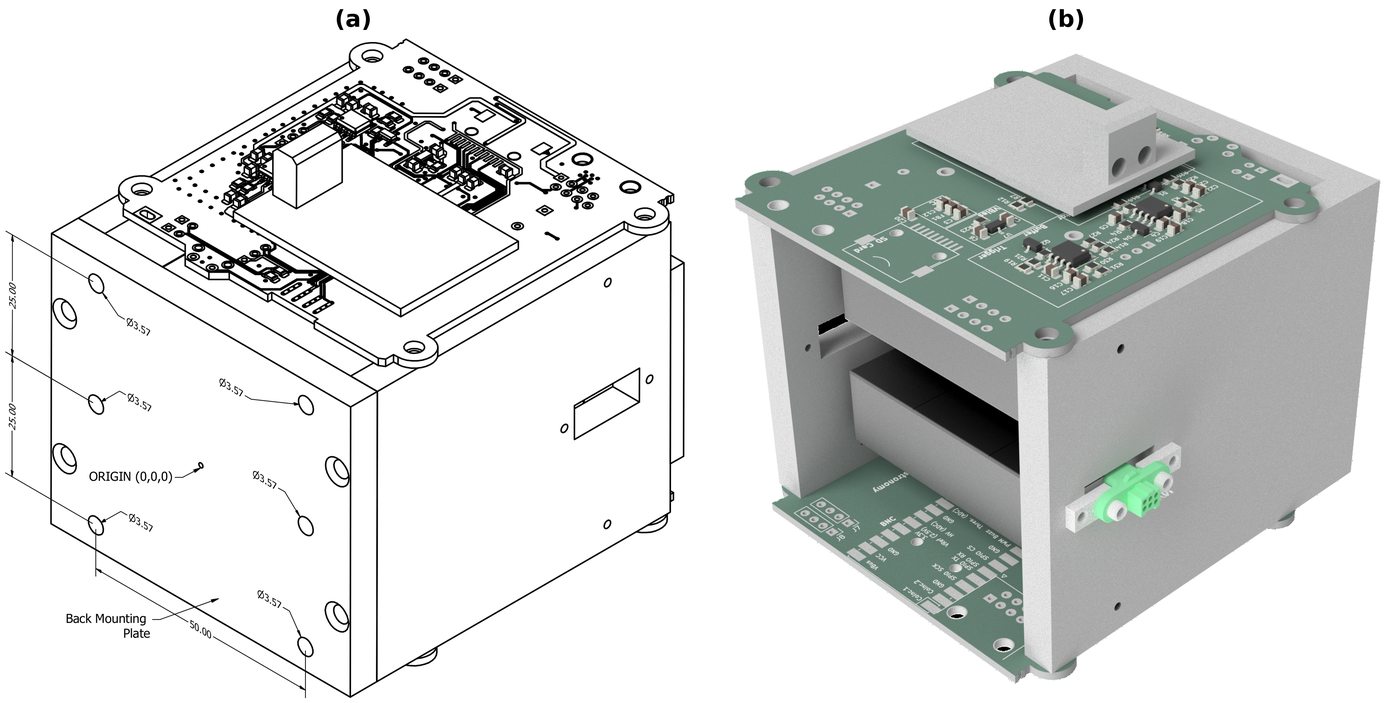}
\caption{The two-detector flight enclosure: (a) isometric technical drawing of the back mounting
plate, showing \O{}3.57\,mm fastener holes on a 25\,mm $\times$ 50\,mm grid; (b) rendered
internal assembly showing the PCB stack, Harwin CAN connector, and microSD slot.}
\label{fig:enclosure}
\end{figure}

The flight configuration packages two complete scintillator detector assemblies into a
machined aluminum case that measures approximately 81$\times$79$\times$79\,mm$^{3}$, shown
in Fig.~\ref{fig:enclosure}. The case is built from two side plates and a back mounting
plate, where the back plate is the single interface between the detector and the satellite;
it carries six \O{}3.57\,mm fastener holes on a 25\,mm $\times$ 50\,mm grid so the case can
be bolted rigidly to the spacecraft structure, with the thermal strap fastening to the same
plate.

\rev{The interface-control model refines the maximum static envelope to 80.85 $\times$ 78.78
$\times$ 78.66\,mm$^{3}$. The six 3.57\,mm clearance holes align with the
spacecraft-provided 6-32 UNC tapped holes, and the payload is mounted with its
detector-stack z-axis toward the spacecraft zenith so that one scintillator remains above
the other for coincidence measurements. The Harwin G125-2240696F5 connector is located on
the back mounting interface at (61.2\,mm, 42.2\,mm, 0\,mm) relative to the payload origin
\cite{ICD}.}

\subsection{CAN Bus Communication}
\label{sec:can}

The flight detector communicates with the host spacecraft through a Controller Area Network
(CAN) interface. Rather than implementing the CAN protocol directly on the RP2040, the
detector uses an external Microchip MCP2515 CAN controller, which interfaces with the
microcontroller over a Serial Peripheral Interface (SPI). The MCP2515 handles the CAN
protocol and connects the detector to the spacecraft CAN bus, allowing detector data to be
transmitted using existing RP2040 hardware peripherals while preserving CPU resources for
data acquisition.

The spacecraft uses a shared CAN bus consisting of a twisted differential wire pair
terminated by 120\,$\Omega$ resistors at each end of the main bus. CAN devices transmit
asynchronously onto the shared bus and use message arbitration to determine which device may
transmit when multiple nodes request access simultaneously. The CosmicWatch detector is
assigned a relatively low message priority compared with other mission-critical spacecraft
devices, so detector data does not interfere with higher-priority spacecraft communications
when the bus is heavily used.

\rev{The flight bus operates at 1\,Mbps. The detector connects as a short, unterminated
branch and therefore does not place a 120\,$\Omega$ resistor across CAN\_H and CAN\_L;
moreover, termination remains at the two physical ends of the spacecraft's main bus. The
payload-side link is bidirectional, although the present interface definition identifies
particle-event transmission as its primary function \cite{ICD}.}

The physical placement of the detector relative to the spacecraft bus was also considered in
the flight-enclosure design. The connection between the detector and the main CAN bus is
limited to approximately 25\,cm to minimize the impedance discontinuities and signal
reflections associated with a long, unterminated branch off the primary bus. A Harwin
connector on the back mounting plate, visible in Fig.~\ref{fig:enclosure}(b), carries this
connection, allowing the detector to mount within the required distance of the spacecraft's
main bus while preserving the mechanical and thermal interfaces described above.

The CAN interface also constrains how detector event data are organized for transmission.
Each event record contains multiple measurements, including the event number, timestamp,
coincidence status, pulse amplitude, accumulated dead time, temperature, pressure,
acceleration, and angular velocity \cite{Axani2025}. While all individual data types
utilized by the detector can be accommodated within a CAN frame, the primary challenge is
efficiently packing these fields into the available CAN data frame. For example,
transmitting several 16-bit quantities around a 64-bit quantity can leave portions of a
frame unused depending on the ordering of the data. The CAN transmission firmware groups
measurements of compatible size together to minimize unused frame space and reduce the
number of CAN messages needed per event.

\rev{The spacecraft allocates approximately 10\,MB of onboard storage per orbit to payload
data, which is equivalent to about 2\,kB/s averaged over an orbit, with one downlink
opportunity per orbit. Efficient frame packing therefore directly increases the number of
events that can be both retained and returned \cite{ICD}.}

\rev{No discrete synchronization or pulse-per-second signal is provided through the
four-wire interface. Event timestamps are generated by the payload clock and can be
associated with spacecraft time when the host receives and processes the data. This does not
change the detector's internal 2.3\,$\mu$s coincidence decision, but it limits the accuracy
of absolute event timing in the spacecraft reference frame \cite{ICD}.}

\rev{During boot, the flight firmware initializes both detector channels and the CAN
interface and begins acquisition within 10\,s. From here, it produces a basic, fixed-format
event log for CAN transmission. Records have a fixed set of numeric fields, including event
number, timestamp, coincidence status, pulse amplitude, and temperature \cite{ICD}.}

\begin{figure}[!htb]
\centering
\includegraphics[width=0.70\columnwidth]{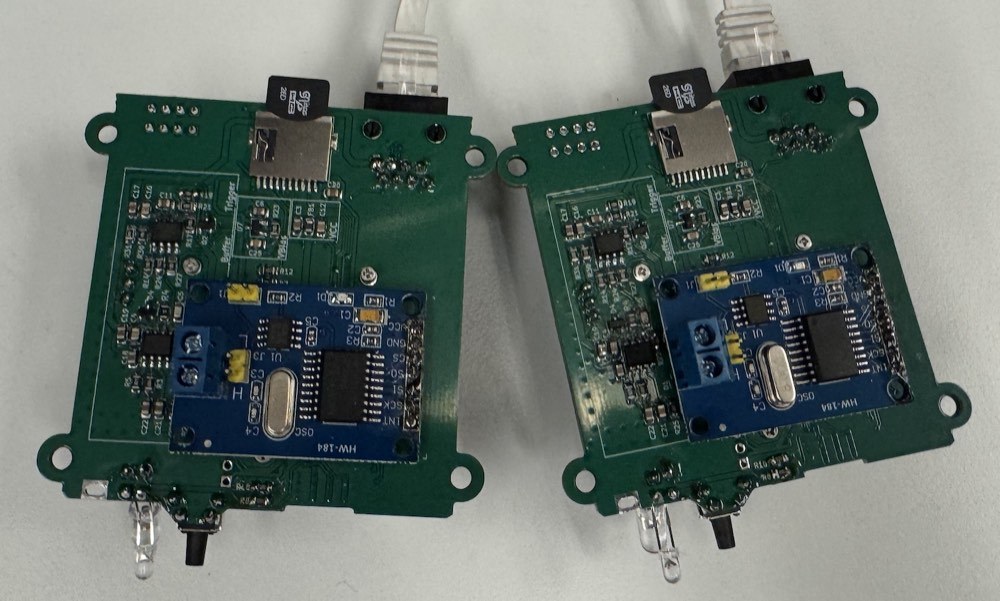}
\caption{\rev{Two populated v3X flight-electronics prototypes with the MCP2515 CAN-controller modules
installed; the revised PCB design most notably includes an additional standoff for improved
vibration support during liftoff. }}
\label{fig:boards}
\end{figure}

\rev{Relative to the desktop v3X, the flight electronics add the external MCP2515
CAN-controller modules and an additional mechanical standoff to support vibration
qualification, and remove the OLED display for flight. Together with the PVT scintillator
and 3M 425 optical-isolation tape described above, these changes distinguish the flight
prototype shown in Fig.~\ref{fig:boards} from the original v3X detector.}

\section{Conclusion}

The CosmicWatch v3X transforms the desktop muon detector into a low-cost, high-rate
instrument with sub-millisecond dead time, native coincidence triggering, and advanced
onboard sensing. It supports measurements from sea-level muon spectra to stratospheric
radiation profiles. Building on this platform, a two-detector flight unit is being developed
for satellite operation. Key features include a passive copper thermal strap bonded with
space-grade epoxy, a machined aluminum flight enclosure, and a CAN bus interface to the host
spacecraft. Room-temperature stress tests confirm the thermal strap's effectiveness.
Upcoming vibration and thermal-vacuum testing, along with repeated CAN-interface test runs,
will further qualify the design before payload delivery. Qualification begins with a
baseline functional test and confirmation that the mass properties are within 5\% of the
declared values, followed by a low-intensity (0.1\,g) 2--128\,Hz sweep in three axes to
confirm that no vibration mode lies below 50\,Hz, static loading to 1.25$\times$, 60\,s of
random vibration per axis, and a repeated sweep that must show less than a 5\% shift in the
primary-mode frequency and under 15\% in its response amplitude; thermal qualification
covers the bakeout described above, 20 acceptance-level thermal cycles, and functional
checks before and after each environmental test \cite{ICD}.

\section*{Acknowledgment}

The authors would like to thank Collin Owens, a graduate researcher in the AxLab group at
the University of Delaware, for his collaboration and guidance on the thermal strap
fabrication, CAN bus communication design in Section \rev{IV-D}, and mission support, and
Masooma Sarfraz, a PhD student in the AxLab group, for her editorial advice and
contributions.

\balance

\end{document}